\documentclass[fleqn,usenatbib]{mnras}

\usepackage{newtxtext,newtxmath}

\def\omegaHI{$\Omega_{\mathrm{HI}}$}
\def\hi{H\,{\sc i}}
\def\deg{$^{\circ}$}
\def\kms{km~s$^{-1}$}
\def\msun{$M_{\astrosun}$}

\usepackage[T1]{fontenc}
\usepackage{ae,aecompl}
\usepackage{wasysym}

\usepackage{graphicx}	
\usepackage{amsmath}	
\usepackage{amssymb}	

\title[Uncertainties of \hi\ Stacking Results]{On the Uncertainties of Results Derived from \hi\ Spectral Line Stacking Experiments}

\author[E. C. Elson et al.]{
E. C. Elson,$^{1,2}$\thanks{E-mail: elson.e.c@gmail.com (ECE)}
A. J. Baker,$^{3}$
$\&$ S. L. Blyth,$^{4}$
\\
$^{1}$Department of Physics $\&$ Astronomy, University of the Western Cape, Robert Sobukwe Rd, Bellville, 7535, South Africa.\\
$^{2}$South African Radio Astronomy Observatory (SARAO), Observatory 7925, South Africa.\\
$^{3}$Department of Physics and Astronomy, Rutgers, The State University of New Jersey, 136 Frelinghuysen Road, Piscataway, NJ 08854-8019, USA.\\
$^{4}$Department of Astronomy, University of Cape Town, Private Bag X3, Rondebosch, 7701, South Africa.}

\date{Accepted XXX. Received YYY; in original form ZZZ}

\pubyear{2019}

\begin{document}
\label{firstpage}
\pagerange{\pageref{firstpage}--\pageref{lastpage}}
\maketitle

\begin{abstract}
We present the results of a set of mock experiments aimed at quantifying the accuracy of results derived from \hi\ spectral line stacking experiments.  We focus on the effects of spatial and spectral aperture sizes and redshift uncertainties on co-added \hi\ spectra, and by implication on the usefulness of results from \hi\ line spectral stacking experiments.  Using large spatial apertures to extract constituent galaxy spectra yields co-added spectra with high levels of contamination and with relatively low signal-to-noise ratios.  These properties are also affected by the size of the spectral aperture as well as \hi\ redshift uncertainties of galaxies.  When redshift uncertainties are high, S/N decreases while the contamination level remains roughly constant. Using small spectral apertures in the presence of large \hi\ redshift uncertainties can yield significant decreases in S/N without the expected decrease in amount of contaminant flux.  Our simulations show that a co-added spectrum rarely yields an accurate measure of the total \hi\ mass of a galaxy sample.  Total mass is generally over/under-estimated for large/small spatial apertures, regardless of spectral aperture size.  Our findings strongly suggest that any co-added \hi\ galaxy spectrum needs to be fully modelled in the ways presented in this paper in order to apply accurate corrections for flux contamination and derive realistic uncertainties in total \hi\ galaxy mass.  Failing to do so will result in unreliable inferences of galaxy and cosmological parameters, such as $\Omega_\mathrm{HI}$.
\end{abstract}

\begin{keywords}
methods: numerical -- radio lines: general -- galaxies: fundamental parameters -- galaxies: evolution
\end{keywords}


\section{Introduction}
Neutral atomic hydrogen (\hi) is  the dominant phase of the interstellar medium at $z=0$.  Constraints on the redshift evolution of the cosmic \hi\ density, $\Omega_{\mathrm{HI}}$, can be used to model the processes by which gas is accreted onto galaxies and subsequently processed and expelled.  The cycle of gas in galaxies (including the time it spends as \hi) is one of the main drivers of their formation and evolution.  

Observations of \hi\ in and around galaxies have traditionally been limited to the nearby Universe.  In the very local Universe, direct detections of \hi\ in galaxies can be used to construct the \hi\ mass function (e.g.,~\citealt{zwaan_HIMF, Martin_2010}) and hence measure $\Omega_{\mathrm{HI}}$.  Owing to the intrinsic faintness of the \hi\ emission line, directly detecting it in galaxies at distances of more than a few hundred megaparsecs requires extremely long integration times.  At high redshifts (z~$\gtrsim1.5$), measurements of \omegaHI\ are derived from observations of damped Lyman~$\alpha$ absorbers (DLAs, e.g.,~\citealt{wolfe_1986, wolfe_1995, lombardi_1996, peroux_2005}).  \citet{sanchez_ramirez_2015}  used a sample of 742 DLAs to measure the continuous redshift evolution of $\Omega_{\mathrm{HI}}$.  By combining their measurements with those at $z=0$, they derive a factor $\sim$~4 decrease in $\Omega_{\mathrm{HI}}$ from $z=5$ to $z=0$.  They do, however, point out the large uncertainties on \omegaHI\ over the intermediate redshift range $0.1\lesssim z \lesssim 1.6$, where it is still poorly constrained by DLAs.  At the low end of this range, stacking of \hi\ data from existing radio telescopes has been used to measure \omegaHI\ (e.g.,~\citealt{chengular_2001, lah_2007, lah_2009, delhaize_2013, rhee_2013, rhee_2016}).  A larger fraction of this redshift range will be probed by means of \hi\ detections of individual galaxies, and spectral line stacking, in the forthcoming Looking At the Distant Universe with the MeerKAT Array  survey (LADUMA, \citealt{LADUMA}), to be carried out on the 64-element MeerKAT array.

Results derived from \hi\  spectral line stacking experiments can be highly uncertain.  For single-dish observations of \hi\ in the nearby Universe as well as interferometric observations of galaxies at higher redshifts, source confusion greatly limits the accuracy with which the average \hi\ content of a sample can be determined by means of stacking.  As an example, for the redshift range $0.04\lesssim z \lesssim 0.13$, \citet{delhaize_2013} estimate any one galaxy in their 2dFGRS sample observed with the Parkes Telescope to be confused, on average, with seven others.  Unfortunately, quantifying such uncertainties is not always possible when considering only the observational data.  \citet{Jones_2015a, Jones_2015b} use the Arecibo Legacy Fast Alfa  (ALFALFA) survey correlation function to develop a computationally inexpensive method of predicting the amount of confused flux in a co-added \hi\ spectrum.  They conclude that stacking in deep SKA-precursor \hi\ surveys will be only mildly affected by source confusion if their target synthesised beam size of 10~arcsec can be achieved.  In \citet{Elson_stacking1}, we presented a set of synthetic data cubes containing model galaxies with realistic spatial and spectral \hi\ distributions based on the catalogue of evaluated galaxy properties from \citet{obresch_2014}.  We used them to carry out several mock \hi\ stacking experiments based on noise-free cubes for low and high-redshift galaxy samples, and consistently found large fractions of contaminant emission due to source confusion in all co-added spectra.  

In addition to quantifying the total amount of confused \hi\ emission in a co-added spectrum, our simulated data cubes can be used to gain further insights into the extent to which  confusion depends on various user-specified extraction parameters, as well as parameters not controlled by the user.  In this work we focus on the sorts of  \hi\ stacking experiments that will be applied to LADUMA \hi\ data.  We do this by  considering a noise-filled version of the high-redshift ($0.7 < z < 0.758$) cube presented in \citet{Elson_stacking1}, which we use to produce a suite of co-added spectra.  The co-adds differ in the ways in which their constituent spectra are generated.  We use spatial apertures of varying size, vary the spectral range over which individual spectra are extracted, and simulate the effects of offsets between optical and \hi\ redshifts.  All of these factors  play a role in determining the amount of contaminant flux in a co-added spectrum, as well as its shape and signal-to-noise ratio.  Our aim in this work is to quantify these various characteristics of a co-added spectrum as functions of the above-mentioned factors.  

The layout of this paper is as follows. In Section~\ref{SyntheticCubes} we present the details of the data cube used for our stacking experiments.  In Section~\ref{Experiments} we describe the various types of co-added spectra we produce in terms of various combinations of aperture sizes and and redshift offsets.  We discuss the various characteristics of our co-added spectra in Section~\ref{Discussion}, and present our conclusions in Section~\ref{conclusions}.  Throughout this work we assume a $\Lambda$CDM cosmology with a Hubble constant $H_0=67.3$~\kms~Mpc$^{-1}$, $\Omega_{\Lambda}=0.685$, and $\Omega_M=0.315$ \citep{Planck_cosmology}.  

\section{Synthetic data cubes}\label{SyntheticCubes}
In this work we make use of the synthetic data products from \citet{Elson_stacking1}, where we presented a set of methods for converting mock galaxy catalogues into realistic data cubes containing \hi\ line emission as well as telescope noise and beam effects.  The cubes are based on the catalogue of evaluated galaxy properties from \citet{obresch_2014}.  For millions of galaxies in the redshift range $z<1.2$, the catalogue presents detailed \hi\ properties as well as auxiliary optical properties.  The catalogue is based on the SKA Simulated Skies semi-analytic simulations, and therefore on the physical models described in \citet{obresch_2009a,obresch_2009b,obresch_2009c}.  These models are able to assign realistic masses and sizes to \hi\ discs, allowing us to evaluate the characteristic properties of their \hi\ emission lines.  In \citet{Elson_stacking1}, we used noise-free cubes to carry out a series of mock \hi\ stacking experiments, in order to quantify the rate of source confusion as a function of spatial resolution.  In this work, we use an almost identical synthetic data cube, but we include noise.  The reader is referred to \citet{Elson_stacking1} for the full details of the properties of the synthetic \hi\ data cubes, as well as the methods used to create them.  

\begin{figure*}
	\includegraphics[width=2\columnwidth]{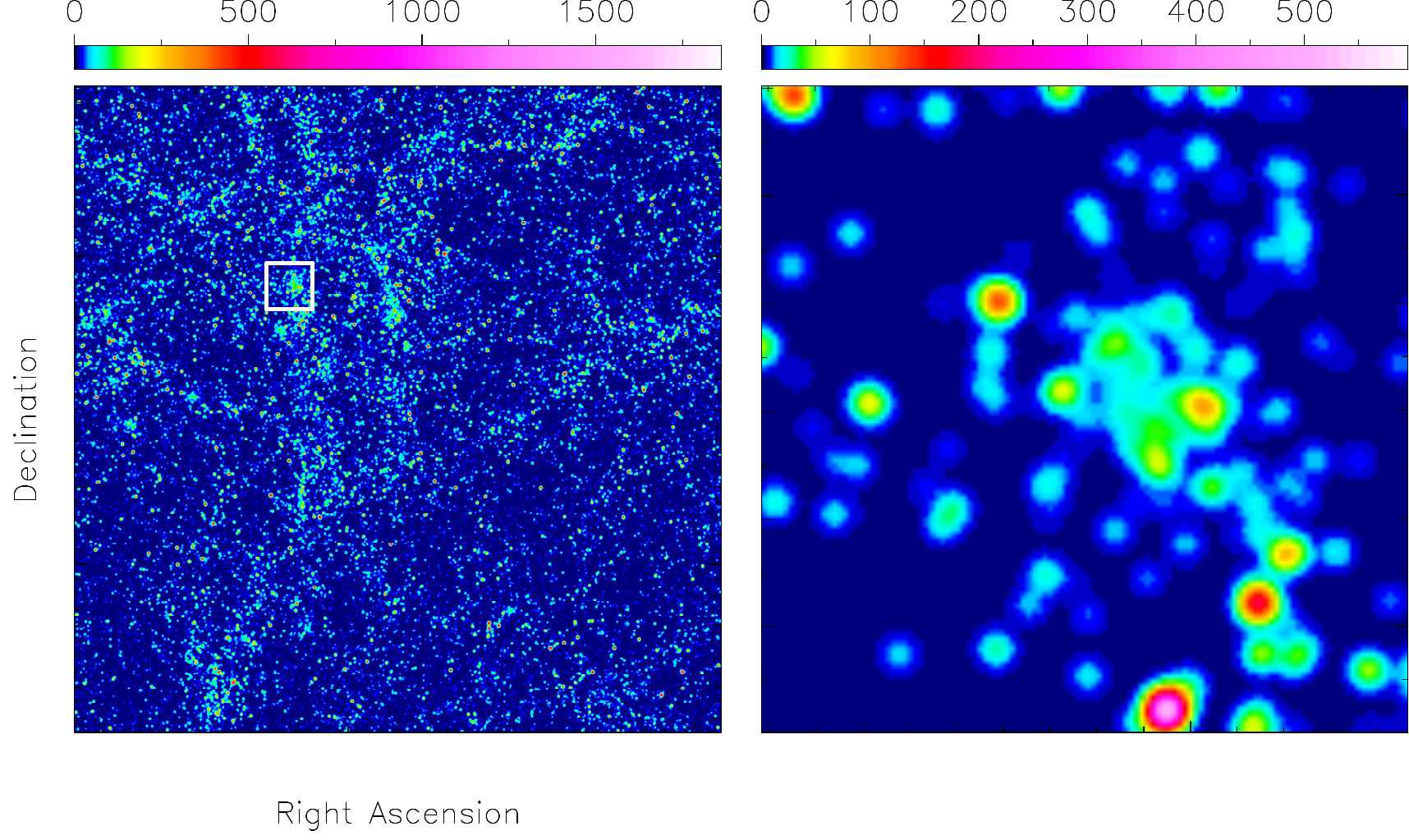}
    \caption{Left panel: Total intensity map for a noise-free version of our synthetic \hi\ data cube, generated by integrating \hi\ emission over the full spectral range of the cube.  The map spans 1.4~\deg$~\times$~1.4~\deg\  and contains \hi\ line emission from 52~662 galaxies in the redshift range 0.7~--~0.758.  The spatial resolution is 15~arcsec.  Right panel: zoom in of the spatial region delimited by the white square in the left panel.  Emission from only 30 consecutive channels is shown, corresponding to a velocity range of $\sim~300$~\kms.  \hi\ flux densities in units of Jy~beam$^{-1}$ to which the colours in each panel correspond are shown in the colour bars.}  
    \label{fig:mom0}
\end{figure*}

We choose to focus on the sort of \hi\ image cube that will typically be produced by  the LADUMA survey.  LADUMA will carry out \hi\ spectral line stacking experiments using data cubes with a frequency-dependent spatial resolution of $\sim~10~-~20$~arcsec.  It was shown in \citet{Elson_stacking1} that \hi\ stacking experiments carried out at $0.7<z<0.758$ will not be dominated by source confusion when the spatial resolution of the data is 18~arcsec.  Rather, $\sim 69$~percent of the flux in a stacked spectrum will come from the target galaxies of interest.  Here we generate a synthetic cube spanning a sky area of 1.4~\deg$~\times$~1.4~\deg\ and a redshift range $0.7< z <0.758$.  This cube contains 52~662 galaxies with a total \hi\ mass of $3.53\times 10^{13}$~\msun.  The distributions of the \hi\  and stellar masses of these galaxies are (statistically) identical to those shown in Fig.~13 of \citet{Elson_stacking1}, to which the reader is referred.  The cube has a spatial resolution of 15~arcsec\footnote{15 arcsec corresponds to  the central frequency, $\nu=820$~MHz, of the redshift range $z=0.7~-~0.758$.}, and spatial and spectral pixel sizes of 3~arcsec and 26~kHz, respectively.  As mentioned above, the main difference between the cubes used in this work and the cubes used in \citet{Elson_stacking1} is the inclusion of noise.  Here we use noise that is Gaussian distributed in each channel, with a standard deviation of 28.8~$\mu$Jy~beam$^{-1}$.  This noise level assumes an estimated system equivalent flux density of 560~Jy for the 64-dish MeerKAT array's UHF receivers at a frequency of $\sim~820$~MHz, for a fiducial integration time of 1000 hours and a channel width of 26~kHz.  Figure~\ref{fig:mom0} shows the total intensity map for the noise-free version of our synthetic \hi\ cube. 

\section{Mock stacking experiments}\label{Experiments}
In this work we focus on quantifying the effects of aperture size, redshift offsets, and spectral extraction ranges of constituent spectra on the characteristics of  co-added \hi\ spectra.  We do not extract the spectra of all 52~662 galaxies in our synthetic cube.  Rather, we consider only those galaxies with stellar mass greater than $10^{10}$~\msun.  This cut results in approximately\footnote{\textcolor{black}{The exact number of individual spectra extracted from the cube depends on the relevant proximities of the galaxies to the edges of the cube.  More specifically, the proximity of the sub-volume (containing the \hi\ line emission of a galaxy) to the edges of the cube is dependent on the choice of sizes of the spatial and spectral apertures used to extract it.}} 2650~-~3000 individual spectra contributing to the various co-added spectra.  

We present our co-added spectra in units of \hi\ mass.  We convert from the native flux density units of the cube to \hi\ mass in a given channel using
\begin{equation}
{M_\mathrm{HI, i}\over M_{\odot}} = {2.36\times 10^5\over 1+z}\left({D_{\mathrm{L}}\over \mathrm{Mpc}}\right)^2\left({S_i\Delta v_i\over \mathrm{Jy \cdot km~s^{-1}}}\right),
\end{equation}
where $M_\mathrm{HI,i}$ is the \hi\ mass in channel $i$, $S_i$ is the flux density in channel $i$, $\Delta v_i$ is the velocity width of channel $i$ \textcolor{black}{in the galaxy rest frame}, $D_{\mathrm{L}}$ is the luminosity distance of a target galaxy, and $z$ is the target galaxy redshift.  From this equation it should be clear that we use the evaluated $D_\mathrm{L}$ and $z$ of \emph{each} galaxy to convert its flux density spectrum into a mass spectrum, rather than using a single measure of  $D_\mathrm{L}$ and $z$ for all stacked galaxies.  \textcolor{black}{In practice, galaxy redshifts are typically obtained from an optical redshift catalogue.  Hence, in spite of the galaxies being un-detected in the HI cube, their spatial and spectral positions within the cube are still known.}

\begin{figure*}
\includegraphics[width=2.1\columnwidth, angle=0]{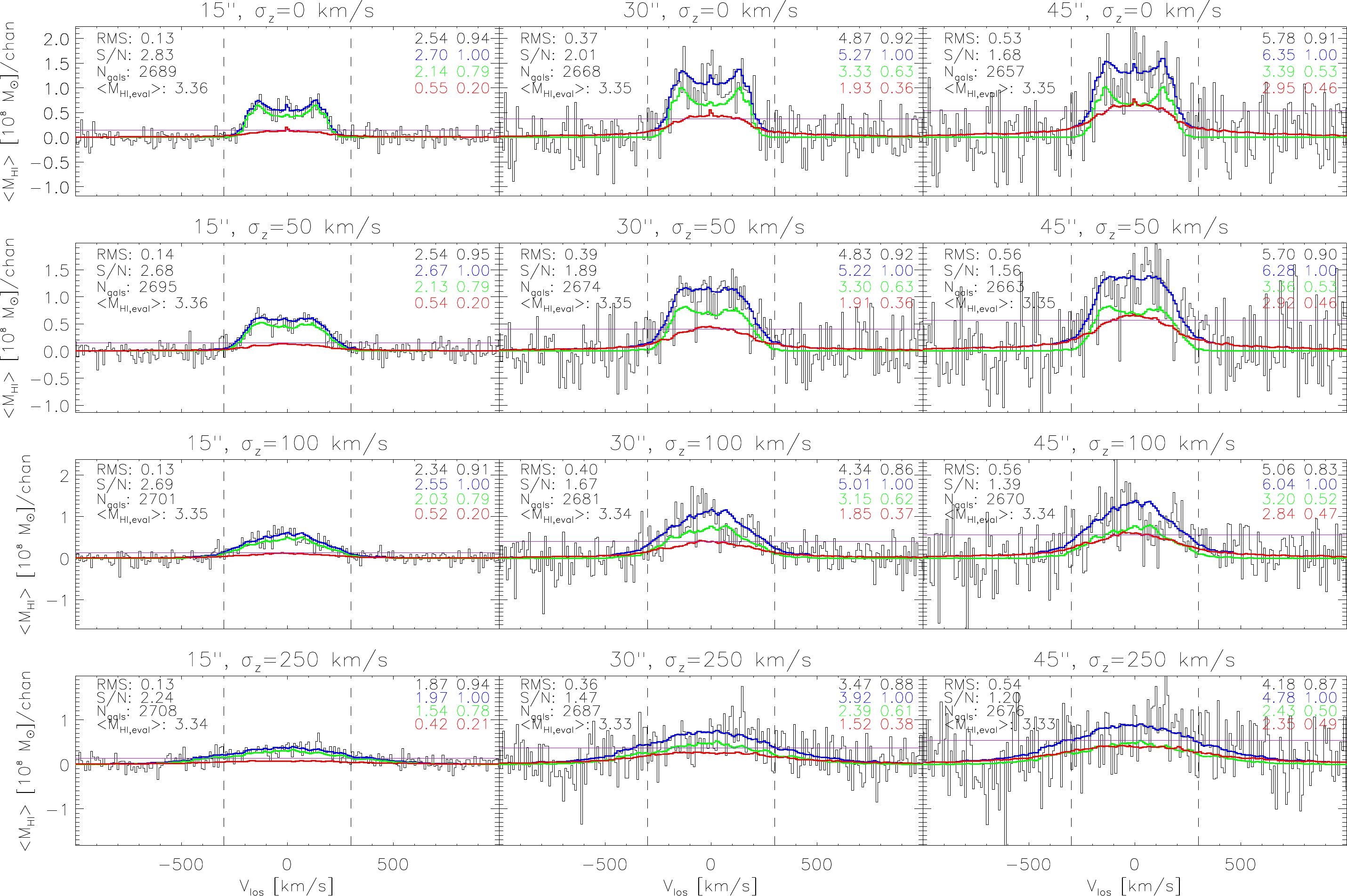}
\caption{Co-added \hi\ spectra based on the long spectra (230 channels) generated for galaxies with $M_*\ge 10^{10}$~\msun\ in our synthetic \hi\ data cube.   From left to right, the constituent  spectra are generated using aperture sizes of 15, 30, and 45~arcsec, respectively.  The co-adds shown in the top row are based on constituent spectra for which the central channel corresponds exactly to the systemic velocity, $V_{sys}$, of the galaxy.  Rows 2, 3 and 4 show co-adds based on constituent spectra for which the central channel corresponds to a velocity equal to $V_{sys}+\delta_z^i$, where $\delta_z^i$ are galaxy-specific velocity offsets drawn from Gaussian distributions with means of 0~\kms and standard deviations of 50, 100, and 250~\kms, respectively.   In each panel, the black histogram represents the galaxy-averaged co-added signal coming from target galaxies, non-target galaxies and noise.  The blue histogram represents the co-added signal coming only from galaxies (targets and non-targets).  The green histogram represents the co-added signal for the target galaxies only.  The red histogram represents the co-added signal from galaxies that are spatially and/or spectrally confused with the target galaxies (i.e., contaminant emission).  The RMS value of the reference spectrum is represented by the horizontal magenta line.  The values shown in the right third of each panel give the mass of a particular co-added mass component in units of $10^9$~\msun\ (left column) and the ratio of that mass to the co-added galaxy mass (right column).  From top to bottom, the values correspond to the co-adds represented by the black, blue, green and red histograms.  Given in the left third of each panel is the RMS variation of the reference spectrum (i.e., the magenta line) in units of \textcolor{black}{$10^{8}$~\msun~} for a single channel, the signal-to-noise ratio for the total co-added signal, the number of galaxies contributing to the co-add, and the true average \hi\ galaxy mass in units of $10^9$~\msun\ as calculated from the evaluated \hi\ masses  in the \citet{obresch_2014} catalogue.  The vertical dashed lines in each panel delimit a spectral range of $\pm 300$~\kms\ about the centre of the co-add.}
\label{fig:longspectra}
\centering
\end{figure*}

\begin{figure*}
\includegraphics[width=2.1\columnwidth, angle=0]{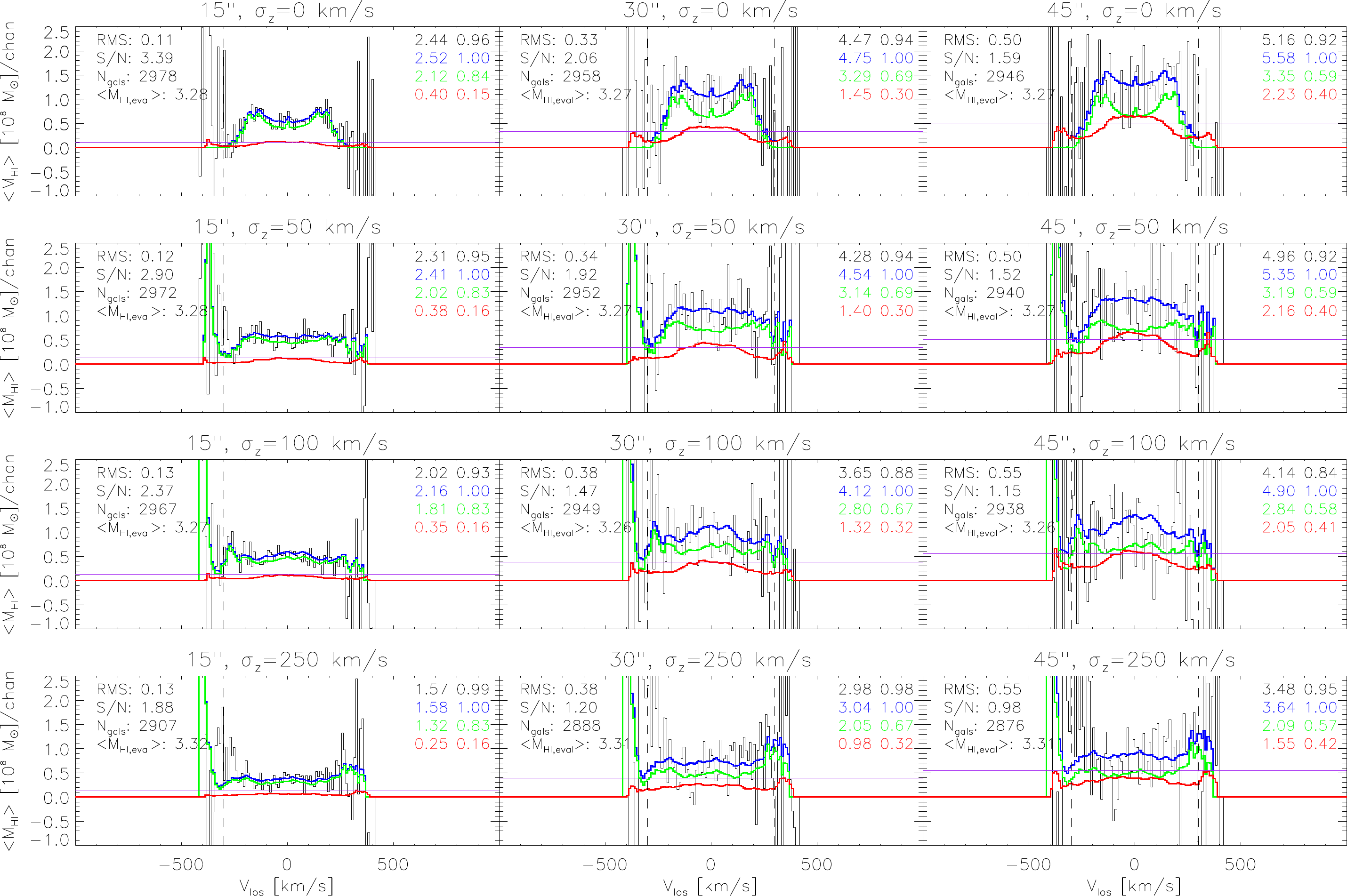}
\caption{Co-added \hi\ spectra based on the mid-length spectra generated for galaxies with $M_*\ge 10^{10}$~\msun\ in our synthetic \hi\ data cube.  Notation is as in Fig.~\ref{fig:longspectra}.  Constituent spectra (such as those in this figure) based on a variable spectral aperture sizes yield co-adds that have a variable number of  galaxies contributing to the flux in each channel.  Because only the most massive galaxies span a velocity range of up to $\sim 600$~\kms, very few galaxies contribute to the channels near the edges of the co-adds.  At these extreme velocities, small number statistics yield highly variable co-added fluxes.}
\label{fig:midspectra}
\centering
\end{figure*}

\begin{figure*}
\includegraphics[width=2.1\columnwidth, angle=0]{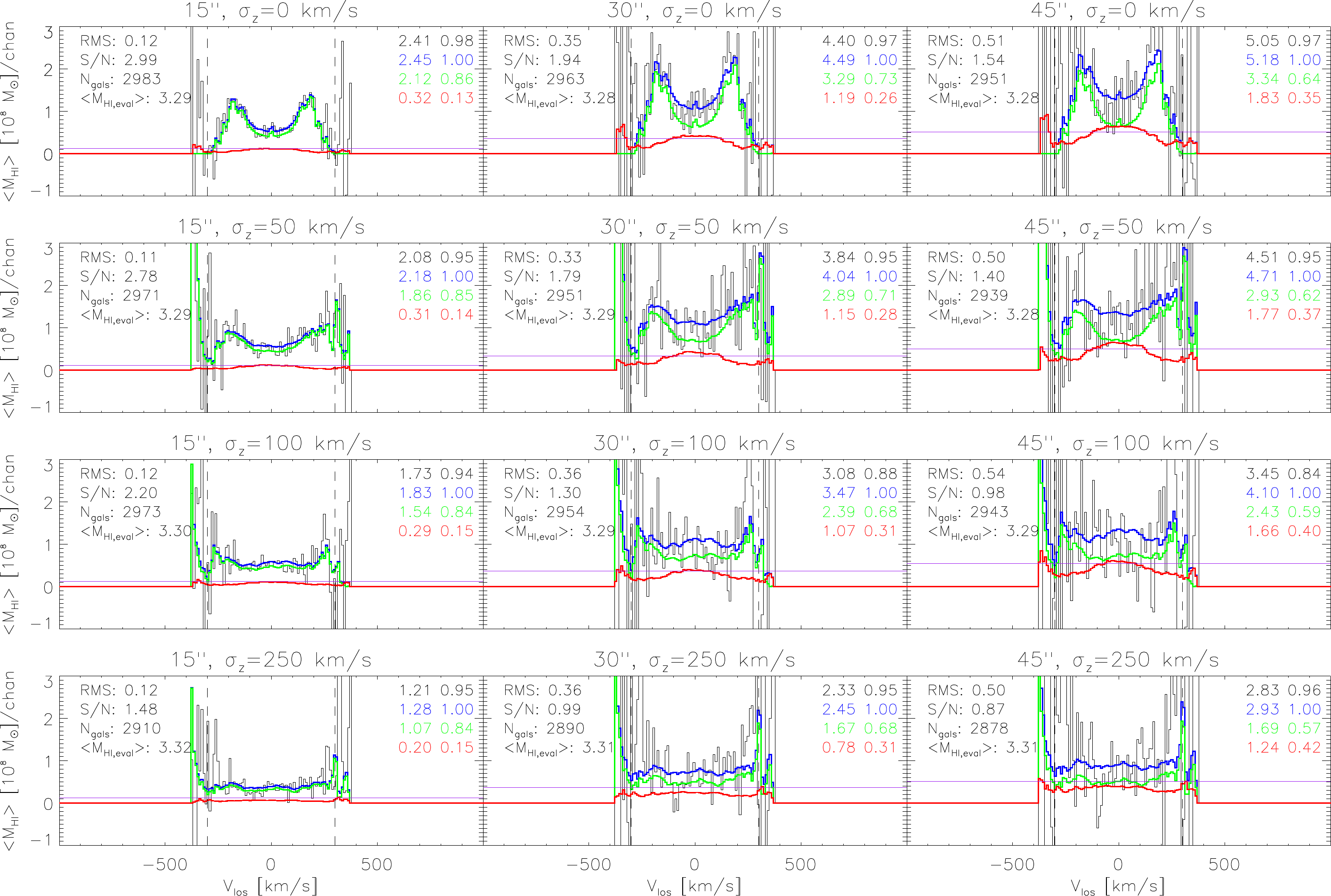}
\caption{Co-added \hi\ spectra based on the short spectra generated for galaxies with $M_*\ge 10^{10}$~\msun\ in our synthetic \hi\ data cube.  Notation is as in Fig.~\ref{fig:longspectra}.  Also see Fig.~\ref{fig:midspectra} caption.}
\label{fig:shortspectra}
\centering
\end{figure*}

In figures~\ref{fig:longspectra}, \ref{fig:midspectra} and \ref{fig:shortspectra},  co-added spectra are decomposed as in \citet{Elson_stacking1}.  For any co-added spectrum, the black histogram represents the galaxy-averaged co-added signal per channel coming from target galaxies, non-target galaxies, and noise.  The blue histogram represents the co-added mass coming from galaxies only, excluding noise.  The green histogram represents the co-added mass coming only from target galaxies, containing zero contaminant signal.  The red histogram represents contaminant mass from nearby  galaxies that are spatially and/or spectrally confused with the target galaxies.  The reader is referred to \citet{Elson_stacking1} for a detailed description of how the mass contributions from target galaxies, nearby neighbours, and distant neighbours are precisely and reliably calculated for a spectrum extracted from the cube.  

Note that throughout this work, we typically refer to the co-added flux coming from target galaxies, non-target galaxies, and noise as the \emph{total} co-added flux or mass.  We do this simply to emphasise the fact that our black histograms in figures~\ref{fig:longspectra}, \ref{fig:midspectra} and \ref{fig:shortspectra} are based on \emph{all} of the flux (including noise) in our simulated cube.  In contrast, the other histograms are based on noise-free versions of our simulated cube.  However, in all of our co-adds, it is always the galaxy-averaged \hi\ mass in a channel that is shown.  In other words, the total amount of co-added flux in a particular channel is always divided by the number of galaxies that contributed flux to that channel. 

\subsection{Co-add variations}
\subsubsection{Aperture size}
An \hi\ spectrum for a single galaxy is usually generated from a small sub-volume spanning several spatial pixels and channels in an \hi\ cube.  The emission in each channel of the sub-volume is spatially integrated to yield flux density as a function of frequency (i.e., a spectrum).  The manner in which a sub-volume is specified for a particular galaxy varies from study to study.  In this work we use the size of the Gaussian beam to specify the spatial extents of the sub-volumes.  We use square spatial apertures with side lengths of 15, 30, and 45~arcsec, corresponding to 1, 2, and 3 half-power widths of the fiducial Gaussian beam\footnote{The width of the fiducial beam is calculated using the central frequency $\nu=820$~MHz of the redshift range $z=0.7~-~0.758$.}.  Given the 3~arcsec spatial pixel scale of our synthetic cube, the apertures have side lengths of 5, 10, and 15 pixels.  

\subsubsection{Spectral range}
\begin{figure}
\includegraphics[width=\columnwidth, angle=0]{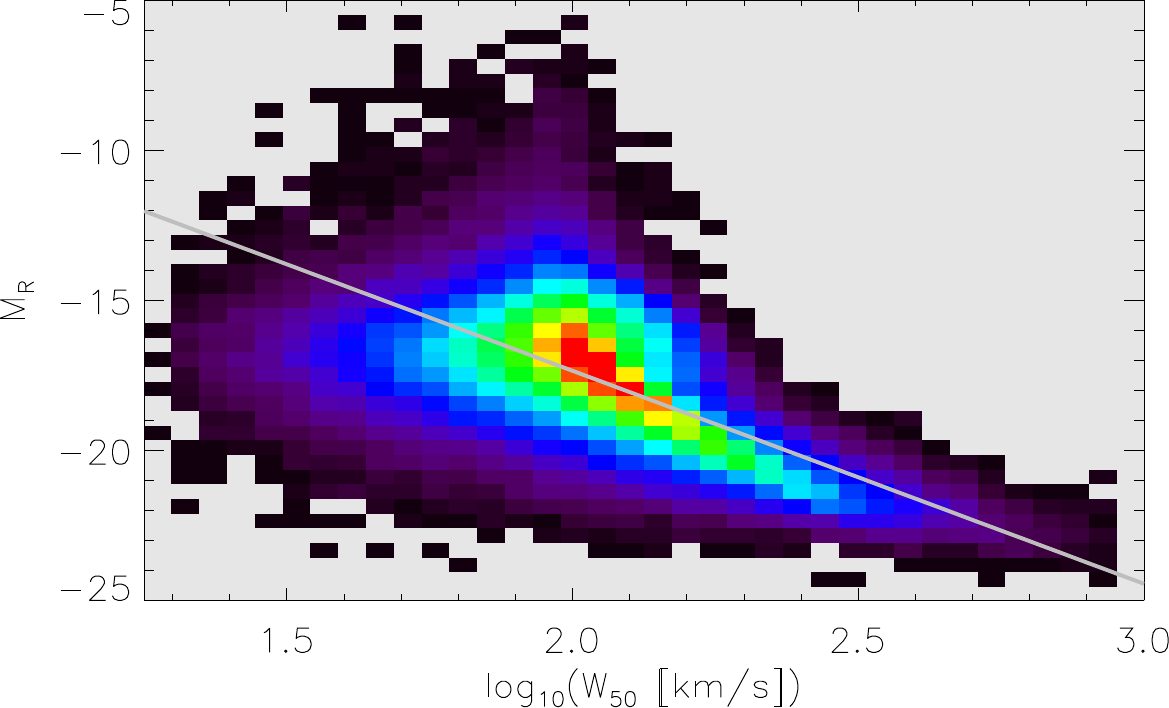}
\caption{Distribution of ($\log_{10}W_{50}, M_{\mathrm{R}}$) ordered pairs for all galaxies in our synthetic cube.  W$_{50}$ measures have been corrected for the evaluated inclinations of the galaxies.  Overlaid as a grey line is the $R$-band Tully Fisher relation for the full \hi\ sample of \citet{verheijen_2001}.}
\label{fig:TFR}
\centering
\end{figure}

We specify the spectral extents of the sub-volumes in three ways.  The first method uses a fixed width of 230 channels, corresponding to a velocity range of $\sim 2184$~\kms.  At the mean redshift of the cube, the frequency width $\Delta\nu=26$~kHz of a channel corresponds to a velocity width $\Delta v = 9.49$~\kms.  These spectra are hereafter referred to as  ``long'' spectra.  The other two methods incorporate a measure of the \hi\ line width of each galaxy.  To more closely mimic real data analysis procedures, rather than using the evaluated \hi\ line width from the \citet{obresch_2014} catalogue (which would not be available a priori for an observed galaxy), we use the evaluated absolute Vega $R$-band magnitude (corrected for intrinsic dust extinction) together with the Tully-Fisher relation \citep{TF_1977} from \citet{verheijen_2001} to calculate each galaxy's \hi\ line width, corrected for the inclination of the disc.  Figure~\ref{fig:TFR} shows the distribution of evaluated ($\log_{10}W_{50}, M_{\mathrm{R}}$) pairs for all galaxies in our synthetic cube, with the \citet{verheijen_2001} relation overlaid.  In order to obtain the line-of-sight \hi\ line width of a galaxy, $W_{50}$ must be scaled by $\sin(i)$, where $i$ is the inclination of the galaxy.  In practice, the inclination can be inferred from the optical morphology of a galaxy.  However, galaxies will not be spatially resolved at high redshifts.  We therefore use the \citet{verheijen_2001} Tully-Fisher relation to extract two $W_\mathrm{50}$-based spectral ranges for each galaxy: one based on the assumption that $i=90$~deg, and the other based on the evaluated inclination of the galaxy.  These spectra are hereafter referred to as the ``mid-length'' and ``short'' spectra, respectively. 

\subsubsection{Redshift uncertainty}
Once the spectral range of a sub-volume is specified, the exact spectral location (i.e., channel) in the \hi\ cube about which to centre the sub-volume must also be specified.  If we assume that the \hi\ redshift of each target galaxy is known with zero uncertainty, then we can place the spectral centre of the sub-volume precisely at the corresponding channel in the \hi\ cube.  In practice, however, an accurate measure of the \hi\ redshift of a distant galaxy is seldom known.  For galaxies that fall below the \hi\ sensitivity threshold of a survey, this is always the case.  It is for this reason that the optical redshift of a galaxy is typically used as a proxy for its \hi\ redshift, which in turn allows for the spectral range of the galaxy to be specified.

However, optical and \hi\ galaxy redshifts are known to be typically offset from one another.  In their study of the impact of redshift uncertainties on spectral line stacking, \citet{Maddox_stacking} compare various sets of optical and \hi\ galaxy redshifts.  They find the differences between optical redshifts from the MPA-JHU catalogue and ALFALFA \hi\ redshifts for 6419 galaxies to be well modelled as the sum of two Gaussians with means of 1.94 and -1.53~\kms, and standard deviations of 11.54 and 35.56~\kms, respectively.  They use these Gaussian offsets to carry out some simple stacking experiments.  They also use offsets as large as 250~\kms, inspired by the  redshift survey of the Great Observatories Origins Deep Survey (GOODS) South field conducted by \citet{balestra_GOODS} at low spectral resolution.  For galaxies with more than one spectrum, \citet{balestra_GOODS} find the accuracy of a single measurement to be $\pm$~255~\kms.  

In this work, we model the uncertainties in \hi\ redshift as being Gaussian distributed about 0~\kms, with standard deviations  $\sigma_{\mathrm{z}}=50$, 100, and 250~\kms.  Given $\delta_z^i$ as the redshift uncertainty of galaxy $i$ (drawn from one of these three distributions) that has a true \hi\ redshift $z^i_{\mathrm{true}}$, the spectral centre of its corresponding sub-volume is placed at  the channel corresponding to $z^i_{\mathrm{true}}+\delta_z^i$.  Note that because the random redshift uncertainties (i.e., $\delta_z^i$) are Gaussian-distributed (with a mean of zero), they can be positive or negative.  We also generate a version of each spectrum that is based on a redshift uncertainty of 0~\kms.  

In total, we produce $3\times 3\times 4 = 36$ co-added spectra in this work.  Figures~\ref{fig:longspectra},~\ref{fig:midspectra}, and \ref{fig:shortspectra} show the co-adds based on the long, mid-length and short spectra, respectively.  In each of the figures, columns 1, 2, and 3 correspond to the co-adds based on aperture sizes of 15, 30, and 45~arcsec, respectively.  Row 1 in each of the figures corresponds to co-adds based on spectra that have redshift uncertainties of 0~\kms, whereas rows 2, 3, and 4 correspond to spectra assuming the Gaussian-distributed redshift uncertainties with standard deviations $\sigma_{\mathrm{z}}=25$, 50, and 100~\kms.

\section{Discussion}\label{Discussion}
The various co-adds shown in figures~\ref{fig:longspectra}, \ref{fig:midspectra}, and ~\ref{fig:shortspectra} contain a wealth of information on the effects that  spatial and spectral aperture sizes and redshift uncertainties have on the characteristics of co-added spectra.  Throughout this work, when we state a co-added mass, we refer specifically to an integral of the average co-added mass per channel over the velocity range $-300$~\kms~-~300~\kms\ about the stack centre.  This is the maximum velocity range that the \hi\ line emission from a galaxy is expected to span.  

\subsection{Purity and S/N}
If the appropriate corrections are not applied, a co-added spectrum will seldom provide an accurate measure of the total \hi\ mass of a galaxy sample.  This limitation is due primarily to source confusion in the \hi\ data cube, but also to the specific methodology used to generate the co-add.

In this work, we use the term "purity" to refer to the fractional contribution of target galaxies to the total mass in a co-added spectrum.  Purity is inversely related to the amount of contaminant flux in a stacked spectrum.   The other very important characteristic of a co-added spectrum is its signal to noise ratio (S/N).  Indeed, the main goal of stacking is to combine many low S/N spectra into a single average spectrum with higher S/N.  In this work we conservatively define $S/N$ for a co-added spectrum as the ratio of the total flux integrated over the central $\pm~300$~\kms\ of the co-add to  $N\times \mathrm{RMS_{co-add}}$, where $\mathrm{RMS_{co-add}}$ is the RMS value of a corresponding reference co-added spectrum and $N$ is the number of channels making up the central $\pm~300$~\kms\ of the co-add.  \textcolor{black}{This conservative definition of $S/N$ does not have its denominator scaling as $\sqrt{N}$, as other expressions for integrated signal to noise typically do.  Rather, our definition specifies the total integrated signal as a function of the maximum possible amount of integrated noise.  $S/N$ ratios presented in this work will therefore be lower than those based on the more standard definition, presented in other studies.}  For each constituent galaxy spectrum extracted from the cube, another sub-volume of the same size is extracted at a position offset by 150~arcsec in both right ascension and declination\footnote{For reference: the typical separation between galaxies in our cube with $M_*\ge 10^{10}$~\msun\ is $\sim 0.55$~degrees.}.  This serves as the reference spectrum for the galaxy.  All of the galaxy reference spectra are stacked to yield the final reference spectrum for a co-add.  The RMS value of the reference spectrum (i.e., $\mathrm{RMS_{co-add}}$) is shown as the horizontal magenta line in all panels of figures \ref{fig:longspectra}, \ref{fig:midspectra}, and \ref{fig:shortspectra}.

\begin{figure}
	\includegraphics[width=0.85\columnwidth]{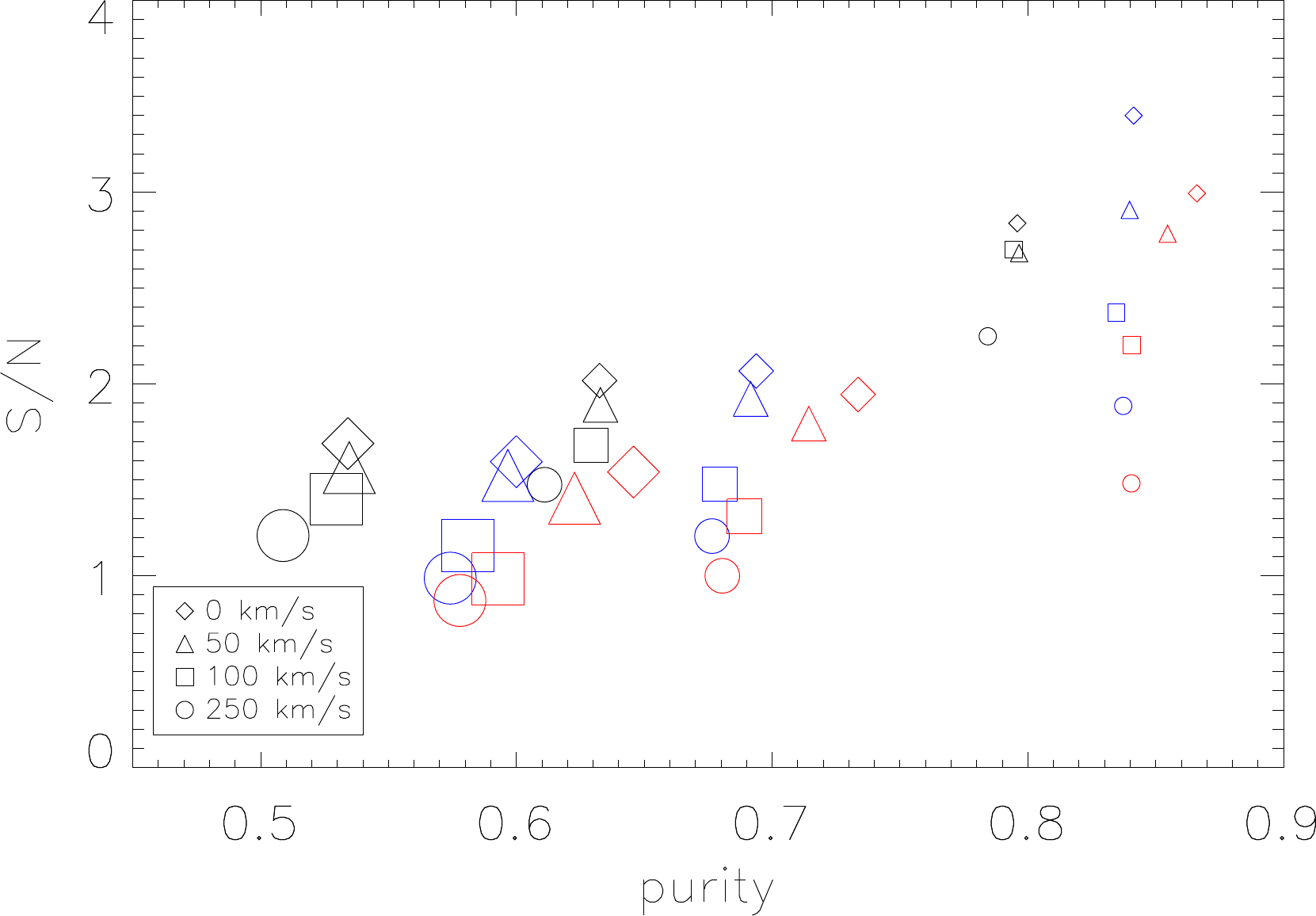}
	\centering
    \caption{Signal-to-noise ratio (S/N) as a function of purity (fractional contribution of target galaxies to total co-added signal) for the co-added spectra shown in figures~\ref{fig:longspectra}, \ref{fig:midspectra}, and \ref{fig:shortspectra}.  Black, blue, and red symbols represent the long, mid-length and short spectra total mass co-adds (i.e., black histograms in figures~\ref{fig:longspectra}, \ref{fig:midspectra}, and \ref{fig:shortspectra}), respectively.  The size of a symbol is proportional to the size of the aperture used to extract the spectra, and the type of symbol represents the standard deviation of the Gaussian-distributed redshift offsets applied to them.  Both purity and S/N are affected by choices of spatial and spectral aperture sizes as well as redshift uncertainties.}
    \label{fig:purity_Vs_SN}
\end{figure}

Figure~\ref{fig:purity_Vs_SN} shows S/N as a function of purity for all 36 of the co-added spectra shown in figures \ref{fig:longspectra}, \ref{fig:midspectra}, and \ref{fig:shortspectra}.  The black, blue, and red symbols represent the co-adds based on  long, mid-length, and short spectra (i.e., figures \ref{fig:longspectra}, \ref{fig:midspectra}, and \ref{fig:shortspectra}).  The colour of a symbol is therefore indicative of the spectral aperture size used to generate the co-add.  The size of any symbol is proportional to the size of the spatial aperture used to generate the co-add (i.e., 15, 30, or 45~arcsec).  Finally, the type (or shape) of a symbol is indicative of the \hi\ redshift uncertainties of the constituent spectra used to generate a particular co-added spectrum.   

Figure~\ref{fig:purity_Vs_SN} clearly shows both purity and S/N to be affected by the choices of spatial and spectral aperture sizes as well as redshift uncertainties.  Furthermore, the two quantities are linked to one another in a roughly linear fashion.  Digging deeper, we see that the first common characteristic shared by all sets of spectra is the manner in which the purity of a spectrum increases with decreasing spatial aperture size.  All co-adds based on 15~arcsec (i.e., 1 beam width) spatial apertures have $\mathrm{purity} \gtrapprox 0.8$.  Notably, for all co-adds, purity does not vary significantly with increasing redshift offset: for given spatial and spectral aperture sizes, purity decreases by approximately 5~percent as redshift uncertainties increase.  A second characteristic shared by all sets of co-added spectra is the slightly counter-intuitive manner in which S/N (like purity) generally \emph{decreases} significantly with increasing spatial aperture size.  Put differently: more co-added mass \emph{does not} result in a higher S/N, at least not according to the way we define S/N in this work.  As expected, S/N also decreases with increasing redshift offsets.  The typical shape of the co-added profile degrades  significantly (becomes more spectrally extended) with increasing redshift offset as well; however, for a given typical redshift offset, the shape varies relatively little with increasing spatial aperture size.  

Figure~~\ref{fig:purity_Vs_SN} also illustrates the undesirable effects on co-added spectra that can arise from using small spectral apertures when redshift uncertainties are high.  Considering only the 12 small symbols in Fig.~\ref{fig:purity_Vs_SN} (corresponding to 15~arcsec apertures), the black symbols represent the co-adds based on long (velocity length: 2184~\kms) constituent spectra.  Clearly, purity and S/N are roughly constant regardless of \hi\ redshift uncertainties.  It is only when \hi\ redshift uncertainties are of the order of 250~\kms\ that S/N noticeably decreases.  However, the situation is different for the co-adds based on mid-length and short constituent spectra (blue and red symbols, respectively).  These co-adds all span a narrow range in purity, yet a large S/N range.  Specifically, S/N decreases as \hi\ redshift uncertainties increase.  This is due to the manner in which the \hi\ emission from a galaxy can be completely ``missed'' if a mid-length or short spectral aperture is placed at the wrong position in the cube (due to large \hi\ redshift uncertainties).  A detailed example of how spectral ranges and redshift offsets conspire to decrease the purity and S/N of constituent spectra is presented in Appendix~\ref{appendixA}.

\hi\ redshift offsets clearly play an important role in setting the optimal spectral apertures sizes to use in an \hi\ stacking experiment.  If \hi\ redshift offsets are known or expected to be $\ge 100$~\kms, there is no benefit in using very short spectral apertures (velocity length: $W_{50}/\sin(i)$) over mid-length spectral apertures (velocity length: $W_{50}$).  This behaviour therefore eliminates the need for estimates of galaxy inclinations  from optical imaging.  However, given the manner in which S/N decreases significantly when \hi\ redshift uncertainties are high ($\ge 100$~\kms), a good strategy for maximising  the S/N of a stacked spectrum is to use long spectral apertures (not based on $W_{50}$ measures), at the expense of lowering the purity of the co-added spectrum.  

Clearly, the optimal combination of spatial and spectral aperture sizes in the presence of non-zero \hi\ redshift offsets needs to  be uniquely determined for a given \hi\ stacking experiment.  Our simulations and the methods presented in this work can be used to make such determinations to a high level of accuracy.

\subsection{Average \hi\ mass}
While purity and S/N are two important characteristics of a co-added spectrum, the other very important property is the level of accuracy with which the average \hi\ galaxy mass can be determined from the co-add.  Relatively few of the co-adds shown in figures~\ref{fig:longspectra}, \ref{fig:midspectra} and \ref{fig:shortspectra} yield an accurate measure of the true (evaluated) average \hi\ galaxy mass, $\langle\mathrm{M_{HI}}\rangle_\mathrm{true}\approx 3.3\times 10^9$~\msun.  

\begin{figure}
	\includegraphics[width=0.85\columnwidth]{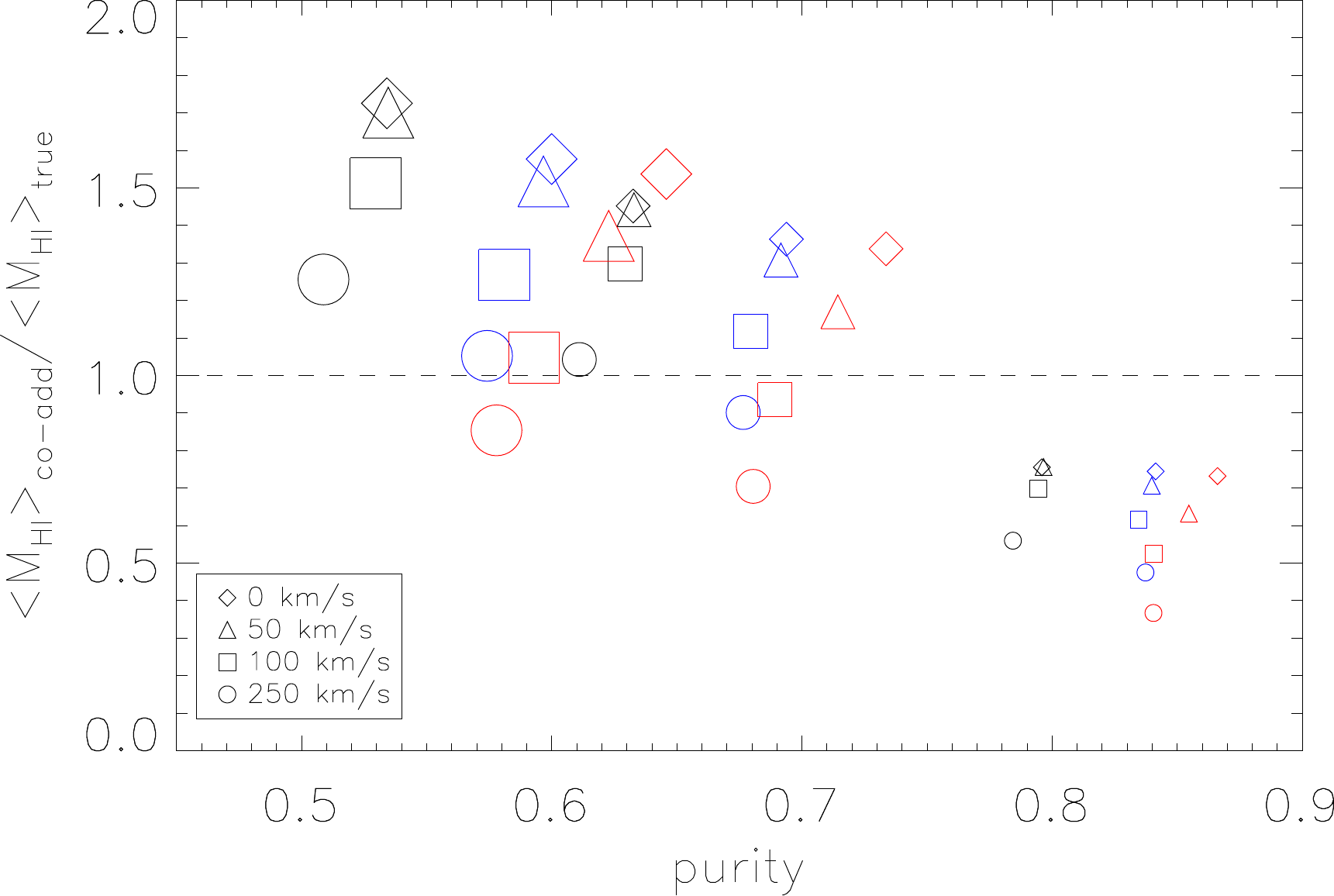}
	\centering
    \caption{Ratio of co-added \hi\ mass to true \hi\ mass, $\langle\mathrm{M_{HI}}\rangle_\mathrm{co-add}/\langle\mathrm{M_{HI}}\rangle_\mathrm{true}$, as a function of purity for the co-added spectra shown in figures~\ref{fig:longspectra}, \ref{fig:midspectra}, and \ref{fig:shortspectra}.  Black, blue, and red symbols represent the long, mid-length and short spectra total mass co-adds (i.e., black histograms in figures~\ref{fig:longspectra}, \ref{fig:midspectra}, and \ref{fig:shortspectra}), respectively.  The size of a symbol is proportional to the size of the aperture used to extract the spectra, and the type of symbol represents the standard deviation of the Gaussian-distributed redshift offsets applied to them.  A co-added spectrum seldom yields an accurate measure of the true \hi\ mass of a galaxy sample.  The size of the spatial aperture used to extract the constituent spectra largely determines the total amount of flux in the corresponding stacked spectrum.}
    \label{fig:mass_variations_specwidth}
\end{figure}

Figure~\ref{fig:mass_variations_specwidth} shows the average \hi\ mass measured from a co-added spectrum, $\langle\mathrm{M_{HI}}\rangle_\mathrm{co-add}$, relative to $\langle\mathrm{M_{HI}}\rangle_\mathrm{true}$ as a function of purity.  Co-adds based on large spatial apertures with \hi\ redshift uncertainties $\leq 50$~\kms\ over-estimate  $\langle\mathrm{M_{HI}}\rangle_\mathrm{true}$ by factors of $\sim 1.4$ to 1.7.  This over-estimate is expected given the high level of contamination (i.e., low purity) of these co-adds.  For \hi\ redshift uncertainties $\ge 100$~\kms, $\langle\mathrm{M_{HI}}\rangle_\mathrm{co-add}/\langle\mathrm{M_{HI}}\rangle_\mathrm{true}$ approaches unity.  However, this convergence is a serendipitous result of the manner in which the large redshift uncertainties degrade the shape and lower the  S/N of the co-added spectrum; it does not suggest that overestimates of $\langle\mathrm{M_{HI}}\rangle_\mathrm{true}$ can be compensated for by large  uncertainties in one's optical redshift catalogue.  

$\langle\mathrm{M_{HI}}\rangle_\mathrm{co-add}/\langle\mathrm{M_{HI}}\rangle_\mathrm{true}$ ratios for co-added spectra based on small (15~arcsec) spatial apertures are all too low by factors of $\sim 0.4$ to 0.8, despite these co-adds having the highest purities.  This combination of high purity and $\langle\mathrm{M_{HI}}\rangle_\mathrm{co-add}/\langle\mathrm{M_{HI}}\rangle_\mathrm{true}<1$ is a result of the manner in which a spatial aperture smaller than three times the half-power width of the resolution element naturally probes only a subset of the spatial area over which the \hi\ flux of a point source\footnote{All high-redshift galaxies observed with MeerKAT will be spatially unresolved, so the  great majority of their flux will be spatially distributed over the area of the synthesised beam.} is spread.  Co-added spectra based on small spatial apertures will never recover all target galaxy flux.  To achieve full recovery, spatial apertures of size three times or more the half-power width of the resolution element are required.  However, as mentioned above, these co-added spectra are highly contaminated, thereby yielding $\langle\mathrm{M_{HI}}\rangle_\mathrm{co-add}/\langle\mathrm{M_{HI}}\rangle_\mathrm{true}>1$.  

Co-added spectra based on 30~arcsec spatial apertures yield $\langle\mathrm{M_{HI}}\rangle_\mathrm{co-add}/\langle\mathrm{M_{HI}}\rangle_\mathrm{true}$  similar to those based on the 45~arcsec spatial apertures.  When \hi\ redshift uncertainties are small ($\le 50$~\kms), $\langle\mathrm{M_{HI}}\rangle_\mathrm{true}$ is over-estimated by factors of $\sim 1.2$ to 1.4.  When \hi\ redshift uncertainties are high ($\ge 100$~\kms), $\langle\mathrm{M_{HI}}\rangle_\mathrm{true}$ is underestimated.

Finally, in all of the co-adds presented in ~\ref{fig:longspectra}, \ref{fig:midspectra} and \ref{fig:shortspectra}, the total co-added mass including noise (i.e., the black histograms) is slightly less that the total co-added noise-free mass (i.e., the blue histograms).  In Appendix~\ref{appendixB}, we discuss how this negative contribution of noise to the total co-added mass is entirely consistent with the Gaussian properties of the noise.

The results presented in this section show it to be impossible to generate a co-added spectrum that has simultaneously high purity, high S/N, and $\langle\mathrm{M_{HI}}\rangle_\mathrm{co-add}/\langle\mathrm{M_{HI}}\rangle_\mathrm{true}\approx 1$.  This result again points to the unavoidable need to fully model an \hi\ stacking experiment in the ways presented in this paper, in order to accurately quantify the characteristics of the co-added spectrum and, very importantly, to apply the necessary corrections to any parameters inferred from it (e.g., average \hi\ galaxy mass, $\Omega_\mathrm{HI}$, etc.).

\section{Conclusions}\label{conclusions}
In this work we have presented the details of a suite of  \hi\ spectral line stacking experiments based on synthetic data cubes that mimic the characteristics of observations that will be carried out as part of the MeerKAT deep \hi\ survey, LADUMA.  We have used various combinations of spatial and spectral aperture sizes as well as \hi\ redshift uncertainties to generate co-added spectra.  We analyse the properties of the co-adds in  light of the specific details of how they were generated. 

All co-added spectra contain a significant amount of contaminant emission due to source confusion.  The amount of contaminant emission rises quickly with increasing spatial aperture size.  Spectral aperture size also affects the contamination level.   \hi\ redshift uncertainties have a relatively small effect on the contamination level, yet significantly impact the signal-to-noise ratio (S/N) of a co-added spectrum. Using small spectral apertures in the presence of large \hi\ redshift uncertainties can lead to a reduction in S/N by a factor $\sim 2$ without lowering the contamination level.  A very clear conclusion drawn from our results is the fact that a co-added spectrum does not produce a reliable measure of the total \hi\ mass of a galaxy sample.  Regardless of spectral aperture size, large/small spatial apertures always over/underestimate the total \hi\ mass.  \hi\ redshift uncertainties need to be carefully considered since they may lower the difference between the true \hi\ mass and the co-added mass, yet in a highly unreliable manner.  Our results suggest that \hi\ spectral line stacking experiments based on LADUMA imaging will yield co-added spectra with a contamination level of up to 20~percent and with S/N (using our new conservative definition) of up to $\sim 3$\footnote{In this work, we define S/N as the ratio of the total co-added flux to co-added noise over the central $\pm 300$~\kms of a stacked profile.  Alternative definitions (e.g., relying on the  peak co-added flux)  will deliver higher S/N but still need to be validated by simulations like those used in this paper.}.

Given that almost all co-added \hi\ spectra contain a non-negligible component of contaminant emission, the appropriate corrections must be made to the total co-added mass in order to retrieve an accurate measure of the contribution from target galaxies alone.  Results based on mock stacking experiments such as ours provide a means of doing so in a  reliable manner.  The scientific impact of \hi\ stacking experiments based on high redshift imaging from forthcoming facilities such as MeerKAT, ASKAP, and ultimately the Square Kilometre Array will be significantly determined by the extent to which we can quantify and correct for the various observational and systematic uncertainties present in quantities derived from co-added spectra.  

\section*{Acknowledgements}

ECE acknowledges the financial assistance of the South African Radio Astronomy Observatory (SARAO) towards this research (www.ska.ac.za).  This work is based on research supported in part by the National Research Foundation of South Africa (Grant Number 115238).  AJB acknowledges support from a Fulbright scholarship during the early stages of this work.  All authors sincerely thank the anonymous referee for insightful feedback that certainly improved the overall quality of this work.  




\bibliographystyle{mnras}



\appendix
\section{Spectral range}\label{appendixA}
When a galaxy spectrum is extracted from a data cube, some estimate of its spectral extent is valuable.  Extracting an unnecessarily long spectrum will lead to the inclusion of unwanted contaminant emission from neighbour galaxies.  A spectral range matching that of the galaxy is expected to minimise the level of contamination in the spectrum.  The Tully-Fisher (TF) relation can be used to estimate the spectral extent of a galaxy.  However, if we are contending with large \hi\ redshift uncertainties, extracting galaxy spectra over a small spectral range can be extremely detrimental to the purity and S/N of the co-add to which they contribute. 

\begin{figure}
	\includegraphics[width=\columnwidth]{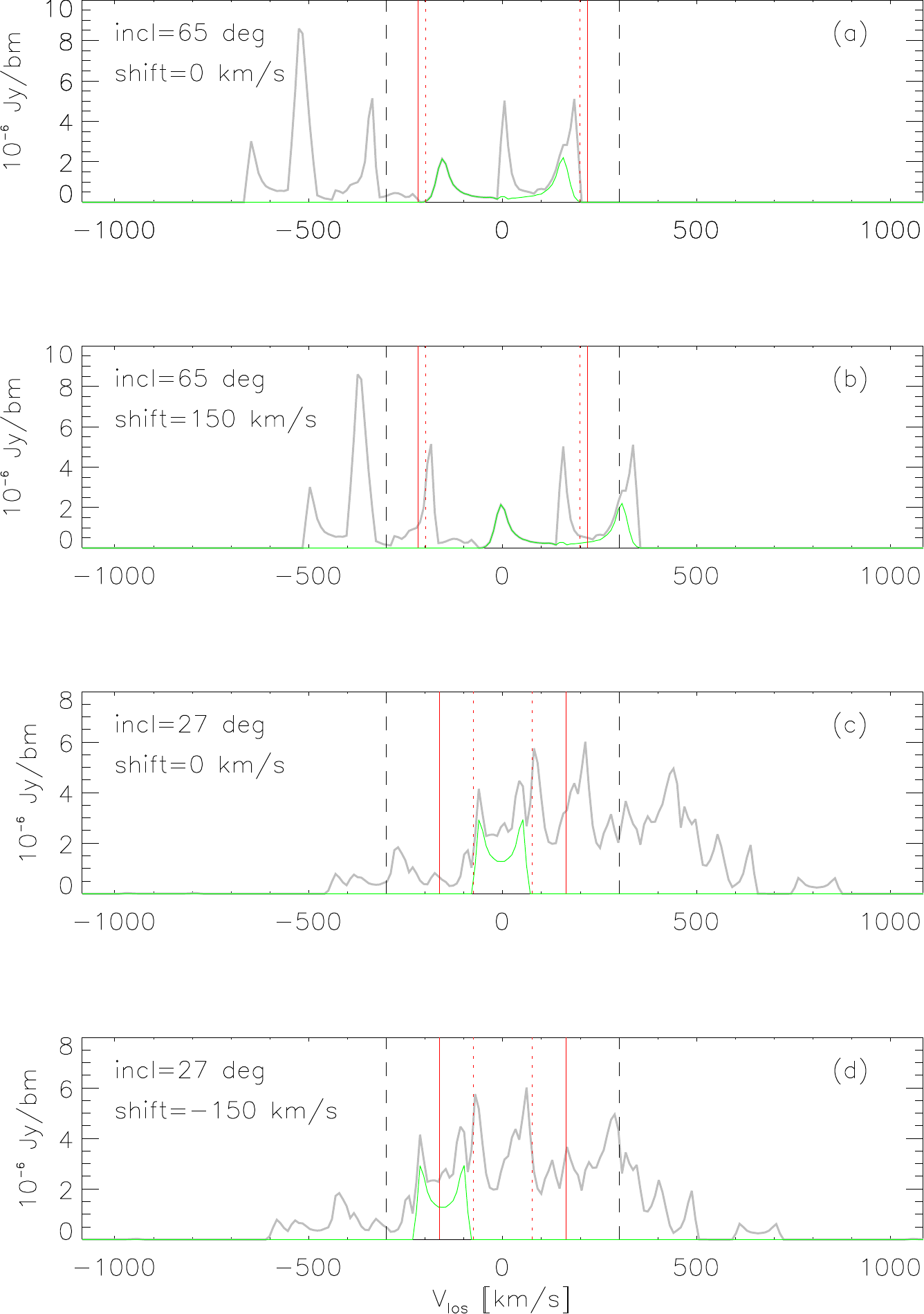}
    \caption{Various spectra associated with each of two different target galaxies.  All spectra are extracted using a spatial aperture size of 30~arcsec.  The top two panels show the spectra of the first galaxy, assuming redshift offsets of 0 and 150~\kms.  The bottom two panels show the spectra of the second galaxy, assuming redshift offsets of 0 and -150~\kms.  In each panel, the green spectrum represents emission from the target galaxy, whereas the grey spectrum is the combined emission from the target and other neighbour galaxies (noise is excluded from this example).  The black-dashed vertical lines delimit a spectral range of $\pm~300$~\kms\ about the centre of the spectrum.  The solid and red-dotted vertical lines represent the TF spectral ranges of the galaxy based on 1) an assumed inclination of 90 degrees,  and 2) the evaluated (true) inclination, respectively.  The actual (evaluated) galaxy inclination is given in the top left of each panel.  A combination of a small spectral extraction range and a large redshift offset can yield an extracted spectrum that completely ``misses''  the emission from a target galaxy of interest, thereby yielding a high level of contamination in the extracted spectrum.}
    \label{fig:laduma_spectral_panel}
\end{figure}

In Fig.~\ref{fig:laduma_spectral_panel}, each panel shows two spectra associated with  one of two different target galaxies,  extracted in several ways from our synthetic \hi\ data cube.  In each panel, the green spectrum represents emission from the target galaxy, whereas the grey spectrum is the combined emission from the target and other neighbour galaxies (noise is excluded from this example). The solid and red-dotted vertical lines represent the TF spectral ranges of the galaxy based on 1) an assumed inclination of 90 degrees,  and 2) the evaluated (true) inclination.

The first (top) panel shows the spectrum of the first galaxy, extracted using a redshift offset of 0~\kms.  It is clear that the two TF-based specifications of the spectral range are very similar to one another and that they both do a good job of excluding the majority of the contaminant emission present within the full (grey) spectrum.  Importantly, they also both neatly capture all of the target galaxy emission.  Panel 2 shows spectra for the same galaxy as in panel 1, this time assuming a redshift offset of 150~\kms.  Due to the large  offset, large fractions of target galaxy emission are shifted rightward of the spectral ranges expected to contain the target galaxy emission.  Worse still is the fact that additional contaminant emission enters both of the TF-based spectral ranges from the left.  The purity of this spectrum is therefore doubly affected, in a detrimental way, by the non-zero redshift offset and the narrow spectral ranges specified by the TF relation.  

The third and fourth panels of Fig.~\ref{fig:laduma_spectral_panel} show the unshifted and shifted versions of the spectra for a second target galaxy from our synthetic cube.  This time the galaxy spans a much smaller spectral range.  The full (grey) spectrum is also highly contaminated by emission from neighbour galaxies.  For a redshift uncertainty of 0~\kms\ (third panel), both TF-based specifications of the target galaxy spectral range do a good job of cutting out almost all of the contaminant emission in the full spectrum.  For the case of a $-150$~\kms\ redshift offset (fourth panel), the situation is entirely different.  The large offset shifts approximately half of the target galaxy emission leftward of the wider TF spectral range (solid red lines), whilst also adding a lot of contamination from the right.  For the narrower TF spectral range based on the evaluated inclination of the galaxy (red-dotted lines), all of the target galaxy emission is shifted leftward beyond its limits, whilst more contamination is again added from the right.  In this case, the combination of a small spectral extraction range and a large redshift uncertainty leads to the galaxy being completely ``missed'', with its extracted spectrum containing only noise.  If redshift offsets are high, the spectral extraction range for each galaxy needs to be kept large enough to ensure that the emission from the target is indeed captured, albeit at the expense of more contamination being introduced.  Simulations such as ours  should be used to find the  optimal method(s) of specifying the spectral range.

\section{Co-added noise}\label{appendixB}
For each of the co-adds shown in Figs.~\ref{fig:longspectra}, \ref{fig:midspectra} and \ref{fig:shortspectra}, the total co-added mass (i.e., black histogram) is less than the co-added galaxy mass (blue histogram).  In other words, the noise in our synthetic cube makes negative contributions to the total co-added mass.  This result, although unexpected,  is indeed statistically correct and is a consequence of the fact that extracting sub-volumes from the full-size cube at the positions of the $M_*\ge 10^{10}$~\msun\ galaxies constitutes a single realisation of the many ways in which a collection of sub-volumes can be extracted.  Repeating such an experiment many times using different sets of constituent spectra does indeed demonstrate that noise on average makes zero contribution to the total co-added mass.

\begin{figure}
	\includegraphics[width=\columnwidth]{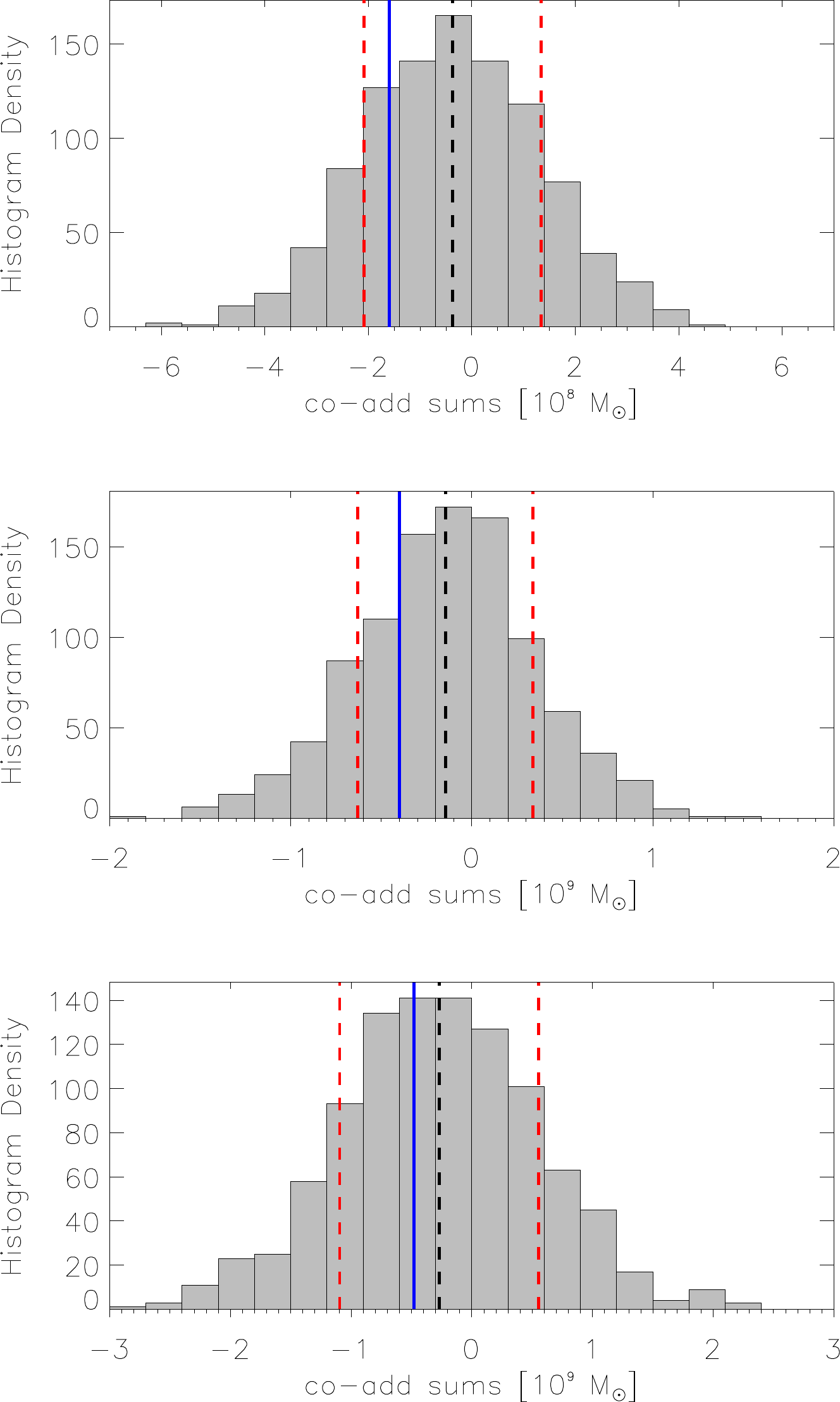}
    \caption{Distributions of summed \hi\ mass for 1000 co-adds, each consisting of 2700 constituent sub-volumes extracted from our smoothed noise cube.  The sub-volumes have pixel dimensions of $N\times N\times 230$, where $N=5, 10, 15$ (top to bottom panels).  The mean, $\mu$, of each distribution is marked by the black-dashed line, and $\mu \pm~1\sigma$ are marked by the red-dashed lines.  The solid blue line in each panel represents total \hi\ mass contributed by noise in our co-added spectra shown in the top row of Fig.~\ref{fig:longspectra}.  In all cases, the (negative) amount of co-added noise is contained well within a single standard deviation of the mean sum of a co-add extracted from the noise cube.}
    \label{fig:N5}
\end{figure}

In this section, we demonstrate this fact by carrying out three experiments in which we use our synthetic noise cube to generate 1000 co-adds each consisting of 2700 constituent spectra.  Each constituent spectrum is based on a sub-volume with pixel dimensions $N\times N \times230$.  For our three experiments, we use $N=5, 10, 15$ to match the spatial pixel dimensions of our long spectra based on 15, 30, and 45~arcsec spatial apertures.  The three panels in Fig.~\ref{fig:N5} show the distributions of the sums of the 1000 co-adds for each of our $N=5, 10, 15$ experiments. The co-add sums are clearly approximately Gaussian distributed with a mean very close to 0 Jy.  In each panel of Fig.~\ref{fig:N5}, the mean co-add sum is marked by the black-dotted vertical line, while the red-dotted vertical lines mark the $\pm~1\sigma$ values.  The (negative) contributions from noise to the total masses in the co-adds shown in Figs.~\ref{fig:longspectra}, \ref{fig:midspectra} and \ref{fig:shortspectra} are entirely consistent with the results presented in Fig.~\ref{fig:N5}.  The most directly comparable cases are the co-adds based on long spectra with redshift offsets of 0~\kms\ (i.e., top row of Fig.~\ref{fig:longspectra}).  For the 15, 30, and 45~arcsec co-adds, the respective noise contributions are $(2.54-2.70)\times 10^9$~\msun~=~$-1.6\times 10^8$~\msun, $(4.87-5.27)\times 10^9$~\msun~=~$-4.0\times 10^8$~\msun, and $(5.78-6.35)\times 10^9$~\msun~=~$-4.8\times 10^8$~\msun.  These contributions are marked by the blue-dashed vertical lines in Fig.~\ref{fig:N5}.  In each case, the contribution is contained well within one standard deviation of the mean sum of the 1000 co-adds generated from our noise cube.  

For each of the distributions shown in Fig.~\ref{fig:N5}, the mean co-add sum (indicated by the black-dashed line) is slightly offset from zero.  However, the smoothed noise cube from which all of the constituent spectra were extracted has a mean voxel value $\mu=-5.06277\times 10^{-9}$~Jy/beam.  This very small, yet negative, mean value yields the observed non-zero mean co-add sums shown in the  panels of Fig.~\ref{fig:N5}.  For example, for the middle panel, $N\times N\times 230\times\mu=-1.16\times 10^{-4}$~Jy/beam, which does indeed convert almost exactly to the mean co-added mass of $=-0.146\times 10^9$~\msun.  This suggests that when given a real cube that has its aggregate line emission dominated by noise (which may also include un-subtracted continuum residuals), co-added fluxes should be appropriately offset in order to account for biases introduced by the noise statistics (specifically a non-zero noise level).

\bsp	
\label{lastpage}
\end{document}